\documentclass[10 pt, conference]{ieeeconf}
\usepackage{balance}
\usepackage[english]{babel}
\IEEEoverridecommandlockouts
\renewcommand{\baselinestretch}{.99}
\usepackage{graphicx}          		
\usepackage{amsmath}           	
\usepackage{amssymb}           	
\usepackage{cite}      		
\usepackage{subfigure}    
\usepackage{theorem}
\newtheorem{theorem}{Theorem}
\newtheorem{lemma}{Lemma}

\newcommand{\qedblack}{\hfill \ensuremath{\blacksquare}}

\title{Linear Approximations to AC Power Flow in Rectangular Coordinates}

\author{Sairaj V. Dhople, Swaroop S. Guggilam, and Yu Christine Chen
\thanks{S. V. Dhople and S. S. Guggilam are with the Department of Electrical and Computer Engineering at the University of Minnesota, Minneapolis, MN 55455, USA. E-mail: \{sdhople,guggi022\}@umn.edu. 
Y. C. Chen is with the Department of Electrical and Computer Engineering, University of British Columbia, Vancouver, British Columbia V6T 1Z4. Email: chen@ece.ubc.ca}}

\begin{document}

\maketitle
\thispagestyle{empty}
\pagestyle{empty}

\begin{abstract}
This paper explores solutions to linearized power-flow equations with bus-voltage phasors represented in rectangular coordinates. The key idea is to solve for complex-valued perturbations around a nominal voltage profile from a set of linear equations that are obtained by neglecting quadratic terms in the original nonlinear power-flow equations. We prove that for lossless networks, the voltage profile where the real part of the perturbation is suppressed satisfies active-power balance in the original nonlinear system of equations. This result motivates the development of approximate solutions that improve over conventional DC power-flow approximations, since the model includes $\mathrm{ZIP}$ loads. For distribution networks that only contain $\mathrm{ZIP}$ loads in addition to a slack bus, we recover a linear relationship between the approximate voltage profile and the constant-current component of the loads and the nodal active- and reactive-power injections.  
\end{abstract}

\section{Introduction}
The \emph{power-flow} problem is fundamental to all aspects of modelling, analysis, operation, and control of transmission and distribution systems. In a nutshell, it amounts to solving for the nodal voltages in the nonlinear active- and reactive-power balance equations that characterize the sinusoidal steady-state behaviour of AC electrical networks.  Iterative numerical methods are ubiquitous in this regard, since multiplicative and trigonometric nonlinearties quash any fledgling hopes of obtaining analytical closed-form solutions. Having said that, linear approximations that yield accurate estimates of nodal voltages have long been recognized to be useful from computational and analysis perspectives~\cite{Stott-2009,Saverio-2015}.

In this paper, we investigate linearizations of the nonlinear power-flow equations with voltages expressed in rectangular coordinates. The main premise is to solve for a (complex-valued) perturbation vector in rectangular form around an appropriately formulated nominal voltage profile. We provide solutions tailored to the constitutional properties of transmission systems (where generators are modelled as $\mathrm{PV}$ buses and loads are modelled as $\mathrm{ZIP}$ buses) and distribution systems (where, in addition to a slack bus, nodes in the network are modelled as $\mathrm{ZIP}$ buses). Indeed, a key assumption that is made from the outset to obtain the linearized model is that the second-order terms in the power-balance expressions are negligible. To investigate the validity of this assumption, we provide a priori computable bounds on the active- and reactive-power balance errors as appropriate. 

Voltage phasors are typically expressed in polar coordinates in most renditions of the power-flow equations. Noteworthy exceptions where rectangular-coordinate representations are leveraged for analytical and computational benefits include: algorithms for optimal power flow~\cite{Torres96optimalpower,Costa-1999,Zhang-2005}, techniques to identify low-voltage solutions~\cite{Overbye-1996}, solving the power-flow equations for ill-conditioned systems~\cite{Iwamoto-1981}, investigating load-flow feasibility~\cite{Makarov-2000}, and state estimation~\cite{Rao-1983,Keyhani-1985}. (This is by no means an exhaustive survey.) Formulating the power-flow equations with voltages expressed in rectangular coordinates affords us to opportunity to offer a fresh perspective on the ubiquitous \emph{DC power-flow} equations~\cite{Wood:1996}. (See also~\cite{Dorfler-DCPowerFlow-2013,Stott-2009} for recent work in this domain.) Particularly, we formally uncover the restrictive assumptions (flat voltage profile, small angle approximations, neglecting shunt loads and current loads, etc.) with which the linearized solution in rectangular coordinates boils down to the classical DC power-flow approximations. Another theoretical contribution of this work is that in lossless transmission systems including $\mathrm{PV}$ and $\mathrm{ZIP}$ buses, we prove that a purely imaginary complex perturbation around a flat-voltage linearization ensures active-power balance in the original nonlinear power-flow equations. Similar insights have recently been reported in~\cite{Ross-2013}, where the author observes that a purely complex perturbation around a flat start ensures balance of the active-power flows in the original nonlinear equations for lossless networks. (This is also alluded to in~\cite{Baosen-2013}.) Leveraging insights from lossless networks, we hypothesize on a voltage profile for transmission networks as an alternative to the DC power-flow equations that incorporates $\mathrm{ZIP}$ loads, and does not assume the slack bus has unit voltage magnitude.

In addition to the transmission-network setting, we also consider distribution networks that are composed of $\mathrm{ZIP}$ buses in addition to a slack bus (that models the secondary side of the step-down transformer at the feeder head). The nominal voltage is chosen to be the one where the constant-power nodal constraints are ignored; this is referred to subsequently as the~\emph{no-load voltage}. The choice of the no-load voltage is intuitively obvious from a circuit-theoretic vantage point, and proves to be algebraically beneficial in that it yields an analytically tractable linear model. We outline graph- and network-theoretic conditions to establish the uniqueness of solutions to this linearized system. For this setting, the real and imaginary components of the voltage perturbation vector can be solved uniquely without any further structural assumptions since the active- and reactive-power injections at all buses are known. Our results for distribution networks extend recent results in~\cite{Bolognani-2015} to include constant-current loads. We also systematically delineate the restrictive set of assumptions under which it is appropriate to presume that voltage magnitudes are strongly coupled to active-power injections and phases are strongly coupled to reactive-power injections. These assumptions underpin a vast body of work on distribution-system operation and control~\cite{Robbins-2013,Zhong-2013,Emiliano-2015}. 
 
The remainder of this manuscript is organized as follows. Section~\ref{sec:Prelim} establishes notation and describes the power-system model. In Sections~\ref{sec:Transmission} and~\ref{sec:Distribution}, we outline solution strategies suited to transmission and distribution networks, respectively. Concluding remarks are provided in Section~\ref{sec:Conclusions}.

\section{Preliminaries and Power-system Model}
\label{sec:Prelim}

In this section, we establish notation, a few pertinent mathematical preliminaries, and describe the power-system model and the linearization used in the remainder of the paper. 

\subsection{Notation and Mathematical Preliminaries}
The matrix transpose will be denoted by $(\cdot)^\mathrm{T}$, complex conjugate by $(\cdot)^*$, real and imaginary parts of a complex number by $\mathrm{Re}\{\cdot\}$ and $\mathrm{Im}\{\cdot\}$, respectively, magnitude of a complex scalar by $|\cdot|$, and $\mathrm{j} := \sqrt{-1}$. 

A diagonal matrix formed with entries of the vector $x$ is denoted by $\mathrm{diag}(x)$;  
$\mathrm{diag}(x/y)$ forms a diagonal matrix with the $\ell$th entry given by $x_\ell/y_\ell$, where $x_\ell$ and $y_\ell$ are the $\ell$th entries of vectors $x$ and $y$, respectively; and $\mathrm{diag}(1/x)$ forms a diagonal matrix with the $\ell$th entry given by $x_\ell^{-1}$. For a matrix $X$, $x_{\ell m}$ returns the entry in the $\ell$ row and $m$ column of $X$. The null space of a matrix $X$ is denoted by $\mathrm{N}(X)$. For a vector $x=[x_1,\dots,x_N]^\mathrm T$, $\cos(x):= [\cos(x_1),\dots,\cos(x_N)]^\mathrm T$ and $\sin(x):= [\sin(x_1),\dots,\sin(x_N)]^\mathrm T$. We will routinely decompose the complex-valued vector $x \in \mathbb{C}^N$ (complex-valued matrix $X \in \mathbb{C}^{N \times N}$) into its real and imaginary parts as follows: $x = x_\mathrm{re} + \mathrm j x_\mathrm{im}$ ($X = X_\mathrm{re} + \mathrm j X_\mathrm{im}$, respectively). 

The spaces of $N\times1$ real-valued and complex-valued vectors are denoted by $\mathbb{R}^N$ and $\mathbb{C}^N$, respectively; $\mathbb{T}^N$ denotes the $N$-dimensional torus. 
The $N \times N$ identity matrix is denoted by $I_{N\times N}$. The $M \times N$ matrices with all zeros and ones are denoted by $\mathbf{0}_{M\times N}$ and $\mathbf{1}_{M\times N}$, respectively; similarly, the $N \times 1$ vectors with all zeros and ones are denoted by $\mathbf{0}_{N}$ and $\mathbf{1}_{N}$, respectively. 

In subsequent developments, we will routinely employ the following norms. The standard $2$-norm of the vector $x \in \mathbb{C}^N$ is denoted by $||x||$, and defined as
\begin{equation}
||x|| := \left( \sum_{\ell=1}^N |x_\ell|^2 \right)^{\frac{1}{2}}.
\label{eq:2norm}
\end{equation}
We will also find the following norm for the matrix $A \in \mathbb{C}^{N \times N}$ useful~\cite{Bolognani-2015}:
\begin{equation}
||A||^\dagger := \max_\ell \left( \sum_{k=1}^N |a_{\ell k}|^2 \right)^{\frac{1}{2}}.
\label{eq:mtxNorm}
\end{equation}

We conclude this section, with two bounds that we will utilize to bound error terms in the remainder of the paper. Let $||x||$ be the 2-norm of a complex-valued vector, $x \in \mathbb{C}^N$, as defined in~\eqref{eq:2norm}, and $||A||^\dagger$ be the matrix norm of a complex-valued matrix, $A \in \mathbb{C}^{N \times N}$, as defined in~\eqref{eq:mtxNorm}.  Then
\begin{align}
||\mathrm{diag} (x) A x|| &\leq ||A||^\dagger ||x||^2, \label{eq:bound1} \\ 
||A x|| &\leq ||A||^\dagger ||x||. \label{eq:bound2}
\end{align}
These bounds can be derived from Lemma A.1 in~\cite{Bolognani-2015}.

\subsection{Power-system Model}
Consider a power system with $N+1$ buses collected in the set $\mathcal{N}$. We model loads as the parallel interconnection of a constant impedance, a constant current, and a constant power component; in the literature, this is commonly referred to as a $\mathrm{ZIP}$ model~\cite{price1993load}. On the other hand, we model generators as $\mathrm{PV}$ buses, i.e., at generator buses the voltage magnitudes and active-power injections are fixed.

Without loss of generality, the slack bus is fixed to be the $N+1$ bus, and its voltage is denoted by $V_\circ \mathrm{e}^{\mathrm j \theta_\circ}$. Let $V = [V_1, \dots, V_{N}]^\mathrm{T} \in \mathbb{C}^N$, where $V_\ell = |V|_\ell \angle \theta_\ell \in \mathbb{C}$ represents the voltage phasor at bus $\ell$. In subsequent developments, we will find it useful to define the vectors $|V| = [|V|_1,\dots,|V|_N]^\mathrm T \in \mathbb{R}^N_{>0}$ and $\theta = [\theta_1,\dots,\theta_N]^\mathrm T \in \mathbb{T}^N$. Given our focus on rectangular coordinates, we will also routinely express $V = V _\mathrm{re} + \mathrm j V_\mathrm{im}$, where $V_\mathrm{re}, V_\mathrm{im} \in \mathbb{R}^N$ denote the real and imaginary components of $V$.

Let $I = [I_1, \dots, I_{N}]^\mathrm{T}$, where $I_\ell \in \mathbb{C}$ denotes the current injected into bus $\ell$.  Kirchhoff's current law for the buses in the power system can be compactly represented in matrix-vector form as follows:
\begin{equation}
\left[ \begin{array}{c} I  \\  I_{N+1} \end{array} \right] = \left[ \begin{array}{c c} Y & \overline{Y}\\ \overline{Y}^{\mathrm T}  & y  \end{array} \right] \left[ \begin{array}{c} V \\ V_\circ \mathrm e^{\mathrm j \theta_\circ} \end{array} \right],
\label{eq:Ii}
\end{equation}
where $V_\circ \mathrm e^{\mathrm j \theta_\circ}$ is the slack-bus voltage, $I_{N+1}$ denotes the current injected into the slack bus, and the entries of the admittance matrix have the following dimensions: $Y \in \mathbb{C}^{N\times N}$, $\overline Y \in \mathbb{C}^{N}$, and $y \in \mathbb{C} \setminus \{0\}$. 

Corresponding to the matrix $Y = G + \mathrm{j}B$, where $ G, B \in \mathbb{R}^{N\times N}$,  we will denote the vector of shunt admittances (that appear on its diagonal) by $Y_\mathrm{sh} \in \mathbb{C}^N$, and write $Y_\mathrm{sh} = G_\mathrm{sh} + \mathrm j B_\mathrm{sh}$, where $G_\mathrm{sh}, B_\mathrm{sh} \in \mathbb{R}^N$. Exploiting the construction of the admittance matrix, we can extract these shunt elements through:
\begin{equation}
Y \mathbf{1}_N + \overline Y = Y_\mathrm{sh} = G_\mathrm{sh} + \mathrm j B_\mathrm{sh}.
\label{eq:Ysh}
\end{equation}
Note that $Y_\mathrm{sh}$ includes both the shunt terms from the transmission-line lumped-element model as well as those originating from the constant-impedance component of the $\mathrm{ZIP}$ load model. 

\noindent \textbf{Remark.} The matrix $Y$ does~\emph{not} correspond to the admittance matrix of a realizable AC electrical circuit. Nonetheless, we prove in Lemma~1 that it is nonsingular by virtue of irreducible diagonal dominance~\cite{Horn:2013} if removal of the slack bus does not affect connectivity of the remaining network.

Denote the vector of complex-power bus injections by $S = [S_1, \dots, S_{N}]^\mathrm T$, where $S_\ell = P_\ell + \mathrm j Q_\ell$. By convention, $P_\ell$ and $Q_\ell$ are positive for generators and negative for loads (these represent the constant power component of the $\mathrm{ZIP}$ load model). Furthermore, let $I_\mathrm{L} = [I_{\mathrm{L}1}, \dots, I_{\mathrm{L}N}]^\mathrm{T}$, where $I_{\mathrm{L}\ell} \in \mathbb{C}$ denotes the current injected into bus $\ell$ due to the constant current component of the $\mathrm{ZIP}$ load at that bus.  Then, using~\eqref{eq:Ii}, complex-power bus injections can be compactly written as
\begin{align}
S  &=  \mathrm{diag}\left(V\right) I^* -\mathrm{diag}\left(V\right) I_\mathrm{L}^* \nonumber \\
   &= \mathrm{diag}\left(V\right) \left(Y^* V^* + \overline{Y}^* V_\circ \mathrm{e}^{-\mathrm j \theta_\circ}   -  I_\mathrm{L}^* \right).
\label{eq:S}
\end{align}

\subsection{Linearized Power-flow Equations} \label{sec:PowerFlow}
Suppose the solution to the power-balance expression in~\eqref{eq:S} is given by $V^\star \in \mathbb{C}^N$:
\begin{equation}
S = \mathrm{diag}(V^\star) \left(Y^* (V^\star)^* + \overline{Y}^* V_\circ \mathrm{e}^{-\mathrm j \theta_\circ}- I_\mathrm{L}^*\right).
\label{eq:pfstar}
\end{equation}
The abounding nonlinearities in~\eqref{eq:pfstar} preclude the possibility of seeking a closed-form solution to $V^\star$ (or even establish existence of solutions). Therefore, we will seek to linearize~\eqref{eq:pfstar} instead. Central to the linearization approach is to express $V^\star = V + \Delta V$, where $V$ is some a priori determined nominal voltage vector,\footnote{We slightly abuse notation with this formulation, since $V^\star$ (and not $V$) satisfies~\eqref{eq:Ii}. Nonetheless, for ease of exposition, we persist with this notation subsequently.} and entries of $\Delta V$ capture perturbations around $V$.  With $V$ appropriately determined (we comment on the choice of $V$ shortly), we need to solve for $\Delta V$ that satisfies~\eqref{eq:pfstar}.  Substituting $V^\star = V + \Delta V$ in~\eqref{eq:pfstar}, we see that $V+ \Delta V$ satisfies:
\begin{equation} \label{eq:SDeltaS}
S = \mathrm{diag}\left(V + \Delta V\right) \left( Y^* (V + \Delta V)^* + \overline{Y}^* V_\circ \mathrm{e}^{-\mathrm j \theta_\circ}- I_\mathrm{L}^*\right).
\end{equation}
Expanding terms in~\eqref{eq:SDeltaS}, we get 
\begin{align} \label{eq:Sinter}
S &= \mathrm{diag}\left(V\right) Y^* V^* + \mathrm{diag}\left(V\right) Y^* \Delta V^*  \nonumber \\
  &+ \mathrm{diag}\left(\Delta V\right) Y^* V^*  + \mathrm{diag}\left(\Delta V\right) Y^* \Delta V^* \nonumber \\
  &+ \mathrm{diag}\left( V\right)  \overline{Y}^* V_\circ \mathrm{e}^{-\mathrm j \theta_\circ} + \mathrm{diag}\left( \Delta V\right)  \overline{Y}^* V_\circ \mathrm{e}^{-\mathrm j \theta_\circ} \nonumber \\
  &- \mathrm{diag}\left(V\right) I_\mathrm{L}^* - \mathrm{diag}\left(\Delta V\right) I_\mathrm{L}^*.
\end{align}
Neglecting the second-order term, $\mathrm{diag}\left(\Delta V\right) Y^* \Delta V^*$, and recognizing that
\begin{align}
&\mathrm{diag}\left(\Delta V\right) Y^* V^* = \mathrm{diag}\left(Y^* V^*\right)\Delta V, \nonumber \\
&\mathrm{diag}\left( \Delta V\right)\overline{Y}^* V_\circ \mathrm{e}^{-\mathrm j \theta_\circ} = V_\circ \mathrm{e}^{-\mathrm j \theta_\circ}\mathrm{diag}\left(\overline{Y}^*\right)\Delta V, \nonumber \\
&\mathrm{diag}\left(\Delta V\right) I_\mathrm{L}^* = \mathrm{diag}\left(I_\mathrm{L}^*\right) \Delta V, \nonumber 
\end{align}
we can reorganize terms in~\eqref{eq:Sinter} to get
\begin{equation} \label{eq:Sinter1}
\Gamma  \Delta V + \Xi \Delta V^* = S + \Pi,
\end{equation}
where $\Gamma \in \mathbb{C}^{N \times N}$, $\Xi \in \mathbb{C}^{N \times N}$, and $\Pi \in \mathbb{C}^N$ are given by
\begin{align} 
\Gamma &= \mathrm{diag}\left(Y^* V^* + \overline{Y}^* V_\circ \mathrm e^{-\mathrm j \theta_\circ} - I_\mathrm{L}^* \right), \label{eq:A}\\
\Xi &= \mathrm{diag}\left(V\right) Y^*, \label{eq:B} \\ 
\Pi &= -\mathrm{diag}\left(V\right) \left(Y^* V^*  + \overline{Y}^* V_\circ \mathrm e^{-\mathrm j \theta_\circ} - I_\mathrm{L}^*\right). \label{eq:C}
\end{align}

With~\eqref{eq:Sinter1} in place, we turn our attention to solving for the bus-voltage perturbation vector $\Delta V$, using which we could recover an approximation to the actual solution $V^\star$. Decomposing all quantities in~\eqref{eq:Sinter1} into their real and imaginary parts, we can write it equivalently as follows:
\begin{equation}
\left[ \begin{array}{c c} \Gamma_\mathrm{re}+\Xi_\mathrm{re} & -\Gamma_\mathrm{im}+\Xi_\mathrm{im} \\ \Gamma_\mathrm{im}+\Xi_\mathrm{im} & \Gamma_\mathrm{re}-\Xi_\mathrm{re} \end{array} \right]
\left[ \begin{array}{c} \Delta V_\mathrm{re} \\ \Delta V_\mathrm{im} \end{array} \right] = \left[ \begin{array}{c} P + \Pi_\mathrm{re}\\ Q + \Pi_\mathrm{im}\end{array} \right],
\label{eq:linear}
\end{equation}
In general, it is not possible to establish the invertibility of the $2N \times 2N$ matrix in~\eqref{eq:linear}. Some special cases do allow us to establish this, and we dwell on these soon. We next outline some possible choices for the nominal voltage, $V$. 

\subsubsection{Flat Voltage} In a non-stressed system, we can assume all bus-voltage magnitudes are close to $1\,\mathrm{p.u.}$ and phase-angle differences are small. In this simplest case, we set $V = \mathbf{1}_N$. We will utilize this approximation in our study of transmission networks in Section~\ref{sec:Transmission}.

\subsubsection{No-load Voltage}
The structure of the expressions in~\eqref{eq:Sinter1} and~\eqref{eq:A}--\eqref{eq:C} suggest that with the following choice: 
\begin{equation} \label{eq:V}
V = Y^{-1}\left(I_\mathrm L - \overline Y V_\circ \mathrm e^{\mathrm j \theta_\circ} \right),
\end{equation}
we get $\Gamma = \mathbf{0}_{N \times N}$, $\Pi = \mathbf{0}_N$, and subsequently recover the following linearized power-flow expressions 
\begin{equation} \label{eq:DeltaVPowerFlow}
\mathrm{diag}\left(V^*\right) Y \Delta V = S^*.
\end{equation}
Notice that $V$ in~\eqref{eq:V} is the non-zero voltage solution recovered when the current injections in the buses (i.e., $YV + \overline Y V_\circ \mathrm{e}^{\mathrm j \theta_\circ} - I_\mathrm{L}$) are zero. Since this corresponds to the non-zero solution to~\eqref{eq:S} when $S = \mathbf{0}_N$, we refer to it as the \emph{no-load voltage}. It turns out that with the choice of the no-load voltage, we can establish the following result on existence and uniqueness of solutions to~\eqref{eq:DeltaVPowerFlow}.

\begin{lemma}
\label{lemma:Yinv}
A unique solution to~\eqref{eq:DeltaVPowerFlow} exists with the choice of the no-load voltage, $V = Y^{-1}\left(I_\mathrm L - \overline Y V_\circ \mathrm e^{\mathrm j \theta_\circ} \right)$, in~\eqref{eq:V} if
\begin{enumerate}
\item[(i)] Removal of the slack bus does not affect network connectivity. 
\item[(ii)] The constant-current components of the $\mathrm{ZIP}$ buses are such that $I_\mathrm L \neq \overline Y V_\circ \mathrm e^{\mathrm j \theta_\circ}$. 
\end{enumerate}
Condition (i)~establishes invertibility of $Y$ and (ii)~ensures that $V =Y^{-1}\left(I_\mathrm L - \overline Y V_\circ \mathrm e^{\mathrm j \theta_\circ} \right) \neq \mathbf{0}_N$. Taken together, they guarantee that the matrix $\mathrm{diag}\left(V^*\right) Y$ is non-singular, and consequently this ensures~\eqref{eq:DeltaVPowerFlow} admits a unique solution.\end{lemma}

\noindent \emph{Proof.} We first dwell on (i).  Since removal of the slack bus does not affect network connectivity, it follows that the undirected graph induced by $Y$ is connected. This is equivalent to stating that $Y$ is irreducible~\cite[Theorem 6.2.24]{Horn:2013}. Furthermore, by construction of the network admittance matrix, it follows that the entries of $Y$ are such that
\begin{equation*}
|y_{\ell \ell}| \geq \sum_{\ell \neq m} |y_{\ell m}|, \quad \forall \, \ell \in \mathcal{N}\setminus\{ N + 1\}.
\end{equation*}
This implies that $Y$ is diagonally dominant. In addition, for the buses that are connected to the slack bus, a set that we denote by $\mathcal{N}_{N+1}$, it follows that 
\begin{equation*}
|y_{\ell \ell}| > \sum_{\ell \neq m} |y_{\ell m}|, \quad \forall \, \ell \in \mathcal{N}_{N+1}.
\end{equation*}
Under these conditions, it follows that $Y$ is irreducibly diagonally dominant~\cite[Definition 6.2.25]{Horn:2013}, from which we conclude that $Y$ is nonsingular~\cite[Corollary 6.2.27]{Horn:2013}. 

Next, consider the statement in (ii). Since (i) ensures invertibility of $Y$, it follows that $\mathrm{N}(Y) = \mathrm{N}(Y^{-1})= \{\mathbf{0}_N\}$. Therefore, ensuring $I_\mathrm L \neq \overline Y V_\circ \mathrm e^{\mathrm j \theta_\circ}$ is sufficient to guarantee that $V = Y^{-1}\left(I_\mathrm L - \overline Y V_\circ \mathrm e^{\mathrm j \theta_\circ} \right) \neq \mathbf{0}_N$.	

Considered together, conditions (i) and (ii) ensure that the matrix $\mathrm{diag}\left(V^*\right) Y$ is nonsingular, hence guaranteeing a unique solution to~\eqref{eq:DeltaVPowerFlow}. 
\qedblack

We will utilize the no-load voltage profile in analyzing distribution networks in Section~\ref{sec:Distribution}.

\subsection{Errors in Complex-, Active-, and Reactive-power Balance}
To gauge the accuracy of linearized solutions outlined subsequently, we will revert to the quadratic term $\mathrm{diag}(\Delta V)Y^* \Delta V^*$ that we neglected in the derivation of~\eqref{eq:Sinter1}. To this end, we denote the \emph{error in complex-power balance} by $S_\mathrm{h.o.t.}$ and express it as
\begin{align} \label{eq:Shot}
S_\mathrm{h.o.t.} &= \mathrm{diag}(\Delta V)Y^* \Delta V^* \\
&=\mathrm{diag}(\Delta V_\mathrm{re} + \mathrm j \Delta V_\mathrm{im}) (G - \mathrm j B) (\Delta V_\mathrm{re} - \mathrm j \Delta V_\mathrm{im}). \nonumber
\end{align}
From above, isolating the real part of $S_\mathrm{h.o.t.}$, we obtain the \emph{error in active-power balance}, which we denote by $P_\mathrm{h.o.t.}$:
\begin{align} 
P_\mathrm{h.o.t.} &= \mathrm{Re}\{S_\mathrm{h.o.t.}\}  \nonumber \\
&=  \mathrm{diag}\left(\Delta V_\mathrm{re}\right) (G \Delta V_\mathrm{re} - B \Delta V_\mathrm{im}) \nonumber \\ 
                  &+  \mathrm{diag}\left(\Delta V_\mathrm{im}\right) (G \Delta V_\mathrm{im} + B \Delta V_\mathrm{re}). \label{eq:Phot}
\end{align}
Similarly, we isolate the imaginary part of $S_\mathrm{h.o.t.}$, to recover the \emph{error in reactive-power balance}, which we denote by $Q_\mathrm{h.o.t.}$:        
\begin{align}          
Q_\mathrm{h.o.t.} &= \mathrm{Im}\{S_\mathrm{h.o.t.}\} \nonumber \\
&= \mathrm{diag}\left(\Delta V_\mathrm{im}\right) (G \Delta V_\mathrm{re} - B \Delta V_\mathrm{im}) \nonumber \\
                  &- \mathrm{diag}\left(\Delta V_\mathrm{re}\right) (G \Delta V_\mathrm{im} + B \Delta V_\mathrm{re}). \label{eq:Qhot}
\end{align}
Note that $S_\mathrm{h.o.t.}$, $P_\mathrm{h.o.t.}$, and $Q_\mathrm{h.o.t.}$ are all functions of $\Delta V = \Delta V_\mathrm{re} + \mathrm j \Delta V_\mathrm{im}$. However, we will drop this functional dependence to simplify notation. 

\section{Approximations Tailored to Transmission Networks} \label{sec:Transmission}

In this section, we exploit the well-known structural property that transmission networks are mostly inductive~\cite{Wood:1996} to postulate on a solution to~\eqref{eq:linear}. We consider the special case where $G = \mathbf{0}_{N \times N}$ in Section~\ref{sec:inductive}, and then examine the classical DC power-flow approximations in Section~\ref{sec:DCPowerFlow}. 

\subsection{Purely Inductive Network \& Flat-voltage Linearization} \label{sec:inductive}
Consider an inductive network for which $G \approx \mathbf{0}_{N\times N}$.  This model is appropriate for transmission networks where the line inductive elements dominate over the resistive ones. In making this assumption, recognize that we have also neglected all shunt conductance terms (i.e., $G_\mathrm{sh} = \mathbf{0}_N$). For such a network, it turns out that with a linearization around the flat-voltage $V = \mathbf{1}_N$, and under a (technical) constraint on the constant-current component of the $\mathrm{ZIP}$ loads, we can demonstrate that the solution to~\eqref{eq:linear} where $\Delta V_\mathrm{re} = \mathbf{0}_{N}$ ensures zero error in the active-power balance. We formally state and prove this next. Before so doing, we establish some notation to ease exposition. From~\eqref{eq:A}--\eqref{eq:C}, for the particular case $G = \mathbf{0}_{N \times N}$, $G_\mathrm{sh} = \mathbf{0}_N$, $V = \mathbf{1}_N$, and $V_\circ \mathrm{e}^{\mathrm j \theta_\circ} = 1$, with the aid of~\eqref{eq:Ysh}, we get
\begin{align} 
\Gamma &= \mathrm{diag}\left(Y^* V^* + \overline{Y}^* V_\circ \mathrm e^{-\mathrm j \theta_\circ} - I_\mathrm{L}^* \right) \nonumber \\
&= \mathrm{diag}\left(Y^* \mathbf{1}_N + \overline{Y}^* - I_\mathrm{L}^*\right) = \mathrm{diag}(Y^*_\mathrm{sh} - I_\mathrm{L}^*) \nonumber \\
&= \mathrm{diag}(-\mathrm j B_\mathrm{sh} - I_\mathrm{L}^*), \nonumber \\
\Xi &= \mathrm{diag}\left(V\right) Y^* = I_{N\times N} Y^* = -\mathrm j B, \nonumber \\
\Pi &= -\mathrm{diag}\left(V\right) \left(Y^* V^*  + \overline{Y}^* V_\circ \mathrm e^{-\mathrm j \theta_\circ} - I_\mathrm{L}^*\right) \nonumber \\
&= -I_{N\times N}\left(Y^* \mathbf{1}_N + \overline{Y}^*- I_\mathrm{L}^*\right) = - \left( Y^*_\mathrm{sh}- I_\mathrm{L}^* \right) \nonumber \\
&= \mathrm j B_\mathrm{sh}  + I_\mathrm{L}^*.
\end{align}
We will find it useful to define
\begin{align} \label{eq:UsefulDefs}
\Phi_\mathrm{re} &:= \Gamma_\mathrm{re} + \Xi_\mathrm{re} = - \mathrm{diag}(I_\mathrm{L,re}), \\
\Phi_\mathrm{im} &:= -\Gamma_\mathrm{im} + \Xi_\mathrm{im} = - (B - \mathrm{diag}(B_\mathrm{sh})) - \mathrm{diag}(I_\mathrm{L,im}). \nonumber
\end{align}
With the definitions in~\eqref{eq:UsefulDefs}, it follows that following linear equations need be solved to determine the voltage-perturbation vector, $\Delta V$:
\begin{equation} \label{eq:maineq1}
\Phi_\mathrm{re} \Delta V_\mathrm{re} + \Phi_\mathrm{im} \Delta V_\mathrm{im} = P + I_\mathrm{L,re}.
\end{equation}
Note that~\eqref{eq:maineq1} is underdetermined, specifically, with $N$ equations and $2N$ unknowns. We next examine the solution where the real component is suppressed, i.e., $\Delta V_\mathrm{re} = \mathbf{0}_N$, and demonstrate that it  satisfies the active-power balance in the original nonlinear power-flow expressions. We formally prove this statement next. 

\begin{theorem}
\label{prop:inductive_main}
Consider an inductive network, $G = \mathbf{0}_{N \times N}$. Suppose that the constant-current component of the $\mathrm{ZIP}$ loads satisfies the following requirements:
\begin{enumerate}
\item For node $\ell \in \mathcal{N}\setminus\{ N + 1\}$, we have
\begin{equation} \label{eq:ILcondition1}
\Big|-\sum_{\ell \neq m} b_{\ell m} - I_{\mathrm{L,im},\ell} \Big| \geq \sum_{\ell \neq m} |b_{\ell m}|. 
\end{equation}
\item For at least one node in the set $\mathcal{N}_{N+1}$ we have 
\begin{equation} \label{eq:ILcondition2}
\Big|-\sum_{\ell \neq m} b_{\ell m} - I_{\mathrm{L,im},\ell} \Big| > \sum_{\ell \neq m} |b_{\ell m}|. 
\end{equation}
\end{enumerate}
It follows that the voltage profile
\begin{equation} \label{eq:ApproxInductive}
\mathbf{1}_N + \Delta V = \mathbf{1}_N + \mathrm j  \Phi_\mathrm{im}^{-1} (P + I_\mathrm{L,re}),
\end{equation}
where $\Phi_\mathrm{im}$ is specified by~\eqref{eq:UsefulDefs} minimizes the active-power-balance error, $||P_\mathrm{h.o.t.}||$, to zero, i.e., 
\begin{align} \label{eq:InductiveResult1}
\mathrm{arg} \min_{\Delta V \in \mathbb{C}^N} ||P_\mathrm{h.o.t.}|| &= \mathrm j \Phi_\mathrm{im}^{-1} (P + I_\mathrm{L,re}) \nonumber \\
\min_{\Delta V \in \mathbb{C}^N} ||P_\mathrm{h.o.t.}|| &= 0.
\end{align}
Furthermore, the voltage profile in~\eqref{eq:ApproxInductive} yields the following upper bound on the reactive-power balance error
\begin{equation} \label{eq:QhotBound}
||Q_\mathrm{h.o.t.}|| \leq ||B||^\dagger ||\Phi_\mathrm{im}^{-1} (P + I_\mathrm{L,re})||^2.
\end{equation}\end{theorem}

\noindent \emph{Proof.} With the conditions in~\eqref{eq:ILcondition1}--\eqref{eq:ILcondition2}, it follows that $\Phi_\mathrm{im}$ is invertible. Consequently, from~\eqref{eq:maineq1}, we can write 
\begin{align} \label{eq:DeltaVispecial}
\Delta V_\mathrm{im} &= \Phi_\mathrm{im}^{-1} (P + I_\mathrm{L,re}) - \Phi_\mathrm{im}^{-1} \Phi_\mathrm{re} \Delta V_\mathrm{re}.
\end{align}
Also, in the setting where $G = \mathbf{0}_{N \times N}$, we see that $P_\mathrm{h.o.t.}$ from~\eqref{eq:Phot} simplifies to the following
\begin{equation} \label{eq:Photspecial}
P_\mathrm{h.o.t.} =  -\mathrm{diag}\left(\Delta V_\mathrm{re}\right) B \Delta V_\mathrm{im} +  \mathrm{diag}\left(\Delta V_\mathrm{im}\right) B \Delta V_\mathrm{re}.
\end{equation}
Substituting for $\Delta V_\mathrm{im}$ from~\eqref{eq:DeltaVispecial} in~\eqref{eq:Photspecial}, we get after a few elementary algebraic manipulations
\begin{align} \label{eq:Temp1}
P_\mathrm{h.o.t.} &= -\mathrm{diag}(B\Phi_\mathrm{im}^{-1}(P+I_\mathrm{L,re})) \Delta V_\mathrm{re} \nonumber \\
							  &\hspace{-0.4in}+ \mathrm{diag}(\Delta V_\mathrm{re}) \Phi_\mathrm{im}^{-1}\Phi_\mathrm{re} \Delta V_\mathrm{re} + \mathrm{diag}(\Phi_\mathrm{im}^{-1}(P+I_\mathrm{L,re})) B \Delta V_\mathrm{re} \nonumber \\
							  &\hspace{-0.4in} - \mathrm{diag}(\Phi_\mathrm{im}^{-1}\Phi_\mathrm{re} \Delta V_\mathrm{re}) B \Delta V_\mathrm{re}.
\end{align}
Taking the $2$-norm on both sides above, applying the triangle inequality, and utilizing~\eqref{eq:bound1}--\eqref{eq:bound2}, we see that
\begin{align}
||P_\mathrm{h.o.t.}|| &\leq ||\mathrm{diag}(B\Phi_\mathrm{im}^{-1}(P+I_\mathrm{L,re}))||^\dagger ||\Delta V_\mathrm{re}|| \nonumber \\
&\hspace{-0.6in}+||\Phi_\mathrm{im}^{-1}\Phi_\mathrm{re}||^\dagger ||\Delta V_\mathrm{re}||^2 + ||\mathrm{diag}(\Phi_\mathrm{im}^{-1}(P+I_\mathrm{L,re})) B||^\dagger ||\Delta V_\mathrm{re}|| \nonumber \\
&\hspace{-0.6in}+ ||\Phi_\mathrm{im}^{-1}\Phi_\mathrm{re} B^{-1}||^\dagger ||B \Delta V_\mathrm{re}||^2.
\end{align}
Taking the minimum on both sides above with respect to $\Delta V_\mathrm{re}$, and noting that $\mathrm{N}(B) = \mathbf{0}_N$ (see Lemma~1), we see that
\begin{align} \label{eq:temp}
&\min_{\Delta V_\mathrm{re} \in \mathbb{R}^N}||P_\mathrm{h.o.t.}|| \nonumber \\
&\leq \min_{\Delta V_\mathrm{re} \in \mathbb{R}^N} \Big(||\mathrm{diag}(B\Phi_\mathrm{im}^{-1}(P+I_\mathrm{L,re}))||^\dagger ||\Delta V_\mathrm{re}|| \nonumber \\
&\hspace{-0.0in}+||\Phi_\mathrm{im}^{-1}\Phi_\mathrm{re}||^\dagger ||\Delta V_\mathrm{re}||^2 + ||\mathrm{diag}(\Phi_\mathrm{im}^{-1}(P+I_\mathrm{L,re})) B||^\dagger ||\Delta V_\mathrm{re}|| \nonumber \\
&\hspace{-0.0in}+ ||\Phi_\mathrm{im}^{-1}\Phi_\mathrm{re} B^{-1}||^\dagger ||B \Delta V_\mathrm{re}||^2\Big) = 0. 
\end{align}
Note that~\eqref{eq:InductiveResult1} then follows straightforwardly from~\eqref{eq:temp}.

Furthermore, from~\eqref{eq:Qhot}, for the case $G = \mathbf{0}_{N \times N}$, we can see that the error in the reactive-power balance is given by
\begin{equation}
Q_\mathrm{h.o.t.} = -\mathrm{diag}(\Delta V_\mathrm{im}) B \Delta V_\mathrm{im} - \mathrm{diag}(\Delta V_\mathrm{re}) B \Delta V_\mathrm{re}.
\end{equation}
With the choice 
\begin{equation}
\Delta V_\mathrm{re} = \mathbf{0}_N, \quad \Delta V_\mathrm{im} = \Phi_\mathrm{im}^{-1}(P + I_\mathrm{L,re}), 
\end{equation}
we can simplify above to get 
\begin{equation}
Q_\mathrm{h.o.t.} = -\mathrm{diag}(\Phi_\mathrm{im}^{-1} (P + I_\mathrm{L,re})) B \Phi_\mathrm{im}^{-1} (P + I_\mathrm{L,re}).
\end{equation}
Applying~\eqref{eq:bound1} (with the choice $x = \Phi_\mathrm{im}^{-1} (P + I_\mathrm{L,re})$ and $A = B$), we recover the upper bound on the reactive-power-balance error in~\eqref{eq:QhotBound}.
\qedblack

\subsection{Connection to the classical DC Power Flow} \label{sec:DCPowerFlow} 
The voltage profile in~\eqref{eq:ApproxInductive} brings to mind the \emph{classical DC power flow} relations~\cite{Wood:1996}:
\begin{equation} \label{eq:DCapprox}
-\left(B - \mathrm{diag}(B_\mathrm{sh}) \right) \theta = P,
\end{equation}
where $B \in \mathbb{R}^{N\times N}$ is the imaginary part of $Y \in \mathbb{C}^{N\times N}$, $B_\mathrm{sh} \in \mathbb{R}^N$ is the vector with shunt susceptance terms, $\theta \in \mathbb{T}^N$ is the vector of bus phases, and $P \in \mathbb{R}^N$ is the vector of active-power bus injections.\footnote{Typically, in the literature, solutions to the DC power flow expressions are written as $-B^{-1}P$. However, with regard to the notation we use in this work, the matrix $B$ includes shunt terms that are excluded in the classical DC power flow expressions.}

We remark that~\eqref{eq:ApproxInductive} extends the classical DC power flow approximation to the case where the network contains $\mathrm{ZIP}$ loads and assumptions of small bus-voltage angle differences are not made. Next, we demonstrate how~\eqref{eq:DCapprox} can be recovered from~\eqref{eq:Sinter1}, and in so doing, illustrate all the restrictive assumptions imposed in deriving~\eqref{eq:DCapprox}.

\begin{lemma}
The DC power flow approximation in~\eqref{eq:DCapprox} can be derived from~\eqref{eq:Sinter1} under the following assumptions and approximations:
\begin{enumerate}
\item a flat-voltage initialization, $V = \mathbf{1}_N$; 
\item disregarding the constant-current loads, $I_\mathrm{L} = \mathbf{0}_N$;  
\item neglecting shunt conductances, $G_\mathrm{sh} = \mathbf{0}_N$; 
\item setting the slack bus voltage to unity, $V_\circ \mathrm{e}^{\mathrm j \theta_\circ} = 1 \angle 0$; and 
\item the small-angle approximation for voltages $\theta \approx \Delta V_\mathrm {im}$.
\end{enumerate}
\end{lemma}

\noindent \emph{Proof.} Let us begin with~\eqref{eq:A}--\eqref{eq:C}. We set $I_\mathrm{L} = \mathbf{0}_N$, and operate under the flat-voltage assumptions $V = \mathbf{1}_N$, $V_\mathrm{o} = 1\,\mathrm{p.u.}$, and $\theta_\mathrm o = 0^\circ$. Then, with the aid of~\eqref{eq:Ysh}, we get
\begin{align} 
\Gamma &= \mathrm{diag}\left(Y^* V^* + \overline{Y}^* V_\circ \mathrm e^{-\mathrm j \theta_\circ} - I_\mathrm{L}^* \right) \nonumber \\
&= \mathrm{diag}\left(Y^* \mathbf{1}_N + \overline{Y}^*\right) = \mathrm{diag}(Y^*_\mathrm{sh}), \nonumber \\
\Xi &= \mathrm{diag}\left(V\right) Y^* = I_{N\times N} Y^*, \nonumber \\
\Pi &= -\mathrm{diag}\left(V\right) \left(Y^* V^*  + \overline{Y}^* V_\circ \mathrm e^{-\mathrm j \theta_\circ} - I_\mathrm{L}^*\right) \nonumber \\
&= -I_{N\times N}\left(Y^* \mathbf{1}_N + \overline{Y}^*\right) = - Y^*_\mathrm{sh}. \nonumber
\end{align}
Substituting $\Gamma$, $\Xi$, and $\Pi$ from above in~\eqref{eq:Sinter1}, we get
\begin{equation}
\mathrm{diag}\left(Y^*_\mathrm{sh} \right) \Delta V + Y^* \Delta V^* = S - Y^*_\mathrm{sh}.
\label{eq:DC1}
\end{equation}
Assuming $\Delta V_\mathrm{re} = \mathbf{0}_N$, we see that $\Delta V = \mathrm{j}\Delta V_\mathrm{im}$. Consequently, it follows that $\Delta V^* = -\Delta V$, and~\eqref{eq:DC1} simplifies to
\begin{equation}
-\left(Y^* - \mathrm{diag}\left(Y^*_\mathrm{sh} \right)\right)\Delta V = S  - Y^*_\mathrm{sh}.
\label{eq:DC2}
\end{equation}
Isolating the real part of~\eqref{eq:DC2}, we obtain
\begin{equation}
-\left(B - \mathrm{diag}(B_\mathrm{sh}) \right) \Delta V_\mathrm{im} = P - G_\mathrm{sh}.
\label{eq:DC3}
\end{equation}
We see that~\eqref{eq:DCapprox} follows from~\eqref{eq:DC3} by neglecting the shunt conductance terms (i.e., setting $G_\mathrm{sh} = \mathbf{0}_N$), and under the small-angle approximation
\begin{equation}
\theta = \angle(V + \Delta V) = \angle(\mathbf{1}_N + \mathrm j \Delta V_\mathrm{im}) \approx \Delta V_\mathrm {im},
\end{equation}
which completes the proof. \qedblack

\noindent \textbf{Remark.} It is straightforward to see that the error in the solution to $\Delta V_\mathrm{im}$ incurred in neglecting the shunt conductance terms is given by $||\left(B - \mathrm{diag}(B_\mathrm{sh}) \right)^{-1} G_\mathrm{sh}||$. Since $G_\mathrm{sh}$ includes shunt conductance terms from the constant-impedance portion of $\mathrm{ZIP}$ loads (in addition to that emerging from the transmission-line lumped-element model), neglecting $G_\mathrm{sh}$ can potentially result in large errors in the estimate of the bus-voltage angles with the conventional DC power flow approximations.

\section{Approximations Tailored to Distribution Networks} \label{sec:Distribution}
In this section, we consider a network that only contains $\mathrm{ZIP}$ buses in addition to the slack bus. This model is appropriate for distribution networks, where the slack bus models the secondary side of the step-down transformer at the feeder head. We begin with the solution for the most general case, and present some special cases next. 

\subsection{Solution for the General Case}
Since we know the active- and reactive-power injections at all buses, and assuming the conditions of Lemma~1 are met, the solution to~\eqref{eq:DeltaVPowerFlow} can be expressed in analytical closed form as follows:
\begin{equation} \label{eq:DeltaVPQ}
\Delta V =  Y^{-1}\mathrm{diag}\left(1/V^*\right) S^*, 
\end{equation}
with $V$ chosen as in~\eqref{eq:V} from before. The linear approximation to the voltage profile in this setting is therefore given by $V + Y^{-1}\mathrm{diag}\left(1/V^*\right) S^*$. Finally, note that the error in complex-power balance induced by this linear approximation can be obtained from~\eqref{eq:Shot}, and is given by
\begin{equation}
S_\mathrm{h.o.t.} = \mathrm{diag}\left(Y^{-1} \mathrm{diag}\left( 1/V^* \right) S^*\right) \mathrm{diag}\left(1 / V\right) S,
\label{eq:Shot_zip}
\end{equation}
Applying~\eqref{eq:bound1} (with the choice $x = Y^{-1} \mathrm{diag}\left( 1/V^* \right) S^*$ and $A = Y^*$), we recover the following upper bound on the complex-power-balance error:
\begin{equation}
||S_\mathrm{h.o.t.}|| \leq ||Y^*||^\dagger ||Y^{-1} \mathrm{diag}\left(1/V^*\right) S^*||^2.
\end{equation}

\subsection{Revisiting Coupling Arguments}
Let us denote $(G + \mathrm j B)^{-1} = R + \mathrm j X$. Then, expanding the terms in~\eqref{eq:DeltaVPQ}, it is straightforward to separately write out the real and imaginary components of $\Delta V$ as follows: 
\begin{align} \label{eq:DeltaVrPQ}
\Delta V_\mathrm{re} &= \left(R \mathrm{diag}\left(\frac{\cos \theta}{|V|}\right) - X \mathrm{diag}\left(\frac{\sin \theta}{|V|}\right) \right) P \nonumber \\
&+ \left(X \mathrm{diag}\left(\frac{\cos \theta}{|V|}\right) + R \mathrm{diag}\left(\frac{\sin \theta}{|V|}\right) \right) Q,
\end{align}
\begin{align} \label{eq:DeltaViPQ}
\Delta V_\mathrm{im} &= \left(X \mathrm{diag}\left(\frac{\cos \theta}{|V|}\right) + R \mathrm{diag}\left(\frac{\sin \theta}{|V|}\right) \right) P \nonumber \\
&- \left(R \mathrm{diag}\left(\frac{\cos \theta}{|V|}\right) - X \mathrm{diag}\left(\frac{\sin \theta}{|V|}\right) \right) Q.
\end{align}
If we suppose that entries of $V_\mathrm{re}$ and $V_\mathrm{im}$ dominate those in $\Delta V_\mathrm{re}$ and $\Delta V_\mathrm{im}$, respectively, then $|V| + \Delta V_\mathrm{re}$ serves as a first-order approximation to the voltage magnitudes  across the distribution network. Similarly, $\theta + \Delta V_\mathrm{im}$ serves as a first-order approximation to the phases  across the distribution network.

In literature pertaining to distribution system operation and control, a common assumption made to simplify analysis is that voltage magnitudes are strongly coupled to active-power injections and phases are strongly coupled to reactive-power injections~\cite{Robbins-2013,Zhong-2013,Emiliano-2015}. Here, with the aid of~\eqref{eq:DeltaVrPQ} and~\eqref{eq:DeltaViPQ}, we investigate the validity of these assumptions, and the exact topological and constitutional requirements to justify them. First, we have to assume $B = \mathbf{0}_{N \times N}$, following which, we have $R = G^{-1}$ and $X = \mathbf{0}_{N \times N}$ in~\eqref{eq:DeltaVrPQ}--\eqref{eq:DeltaViPQ}. (Note that setting $B = \mathbf{0}_{N\times N}$ neglects the line reactances as well as inductive shunt loads.) We also need to suppose $\cos \theta = \mathbf{1}_N$ and $\sin \theta = \mathbf{0}_N$. Under these assumptions, we get    
\begin{equation*} \label{eq:DeltaVriPQspecial}
\Delta V_\mathrm{re} = G^{-1} \mathrm{diag}\left(\frac{1}{|V|}\right) P, \Delta V_\mathrm{im} = - G^{-1} \mathrm{diag}\left(\frac{1}{|V|}\right)Q.
\end{equation*}
With these restrictive set of assumptions, we see that $|V| + \Delta V_\mathrm{re}$, the first-order approximation to the voltage magnitudes, is only a function of the active-power injections. Similarly, $\theta + \Delta V_\mathrm{im}$, the first-order approximation to the voltage phases across the distribution network, is only a function of the reactive-power injections.

\subsection{Recovering Results in~\cite{Bolognani-2015} as a Special Case}
Consider the special case in which constant-current components in the $\mathrm{ZIP}$ loads are neglected, i.e., $I_\mathrm{L} = \mathbf{0}_N$ in~\eqref{eq:V}. In this case, from~\eqref{eq:V}, the choice of $V$ simplifies to
\begin{equation}
V = V_\circ \mathrm{e}^{\mathrm j \theta_\circ} w,
\label{eq:Vzip}
\end{equation}
where $w = -Y^{-1}\overline{Y}$ (adopting the notation in~\cite{Bolognani-2015}).  Using the choice of $V$ in~\eqref{eq:Vzip}, the solution to~\eqref{eq:DeltaVPowerFlow} becomes
\begin{equation*}
\Delta V = Y^{-1}\mathrm{diag}\left(\frac{1}{V^*}\right) S^* = \frac{\mathrm{e}^{\mathrm j \theta_\circ}}{V_\circ} Y^{-1}\mathrm{diag}\left(\frac{1}{w^*}\right) S^*.
\end{equation*}
This suggests the following linearized voltage profile 
\begin{equation}
V_\circ \mathrm{e}^{\mathrm j \theta_\circ} \left(w + \frac{1}{V^2_\circ} Y^{-1} \mathrm{diag}\left(\frac{1}{w^*}\right) S^* \right)
\label{eq:PQvoltage}
\end{equation}
to be a good first-order approximation for voltages in the distribution network. We mention that this result matches that in~\cite{Bolognani-2015}, which was derived with fixed-point arguments. Finally, we note that with this voltage profile, it follows that
\begin{equation}
||S_\mathrm{h.o.t.}|| \leq \frac{1}{V_\circ^2} ||Y^*||^\dagger ||Y^{-1} \mathrm{diag}\left(1/w^*\right) S^*||^2.
\end{equation}

\section{Concluding Remarks} \label{sec:Conclusions}
This paper examined the classical power-flow expressions written in a compact matrix-vector form and with the nodal voltages expressed in rectangular coordinates. We sought solutions to the nodal voltages in the form of a perturbation around a nominal voltage. For lossless transmission networks, with the usual flat-voltage linearization, we proved that the perturbation where the real component is suppressed yields a voltage profile that satisfies active-power balance in the original nonlinear power-flow expressions. For distribution networks, we demonstrated the analytical convenience of a no-load voltage linearization, and extended recent results on linear voltage approximations to include $\mathrm{ZIP}$ loads.

\balance

\bibliographystyle{IEEEtran}
\bibliography{Bibliography/CC_bibliography}

\end{document}